\documentclass[12pt]{article}
\usepackage{epsfig}
\usepackage{amsfonts}
\usepackage{amscd}
\usepackage{latexsym}
\usepackage{amsmath,amssymb}
\usepackage{verbatim}
\usepackage{setspace}
\usepackage{color}
\usepackage{fancyhdr}
\usepackage{cite}
\usepackage{hyperref}

\usepackage[textheight=9in, textwidth=6.5in, letterpaper]{geometry}

\numberwithin{equation}{section}

\def\e{{\epsilon}}
\def\cs{{\cal S}}

 \def\p{\partial}
 \def\bz{{\bar z}}
 
\def\0{{(0)}}
\def\1{{(1)}}
\def\2{{(2)}}

\def\<{\langle }
\def\>{\rangle }

\numberwithin{equation}{section}

\def\e{{\epsilon}}
\def\cs{{\cal S}}

 \def\p{\partial}
 \def\bz{{\bar z}}
 
\def\0{{(0)}}
\def\1{{(1)}}
\def\2{{(2)}}

\def\<{\langle }
\def\>{\rangle }

\newcommand{\bea}{\begin{eqnarray}}
\newcommand{\eea}{\end{eqnarray}}
\newcommand{\be}{\begin{equation}}
\newcommand{\ee}{\end{equation}}
\newcommand{\ba}{\begin{align}}
\newcommand{\ea}{\end{align}}
\def\be{\begin{equation}}
\def\ee{\end{equation}}
\def\beq{\be\begin{array}{c}}
\def\eeq{\end{array}\ee}

\renewcommand{\epsilon}{\varepsilon}

   \makeatletter
  \let\over=\@@over \let\overwithdelims=\@@overwithdelims
  \let\atop=\@@atop \let\atopwithdelims=\@@atopwithdelims
  \let\above=\@@above \let\abovewithdelims=\@@abovewithdelims
\renewcommand\section{\@startsection {section}{1}{\z@}
                                   {-3.5ex \@plus -1ex \@minus -.2ex}
                                   {2.3ex \@plus.2ex}
                                   {\normalfont\large\bfseries}}

\renewcommand\subsection{\@startsection{subsection}{2}{\z@}
                                     {-3.25ex\@plus -1ex \@minus -.2ex}
                                     {1.5ex \@plus .2ex}
                                     {\normalfont\bfseries}}

\begin{document}

\begin{titlepage}
~~~~~~~~~~~~~~~~~~~~~~~~~~~~~~~~~~~~~~~~~~~~~~~~~~~~~~~~~~~~~~~~~~~~~~~~~

\unitlength = 1mm
\ \\
\vskip 2cm
\begin{center}
{ \LARGE {\textsc{Asymptotic Symmetries and Electromagnetic Memory}}}

\vspace{0.8cm}
Sabrina Pasterski
\vspace{1cm}

{\it  Center for the Fundamental Laws of Nature, Harvard University,\\
Cambridge, MA 02138, USA}

\begin{abstract}
Recent investigations into asymptotic symmetries of gauge theory and gravity have illuminated connections between gauge field zero-mode sectors, the corresponding soft factors, and their classically observable counterparts -- so called ``memories."   Here we complete this triad for the case of large $U(1)$ gauge symmetries at null infinity.
\end{abstract}

\vspace{1.0cm}

\end{center}

\end{titlepage}

\def\vx{{\vec x}}
\def\p{\partial}
\def\po{$\cal P_O$}

\pagenumbering{arabic}

\tableofcontents
\section{Introduction}
Recent investigations into asymptotic symmetries of gauge theory and gravity have illuminated connections between gauge field zero-mode sectors, the corresponding soft factors, and their classically observable counterparts called ``memories."  The connections between these concepts can be illustrated by the following triangle:

\begin{figure}[ht!]
\begin{center}
\includegraphics[width=15cm,clip=true,trim=70 0 20 0]{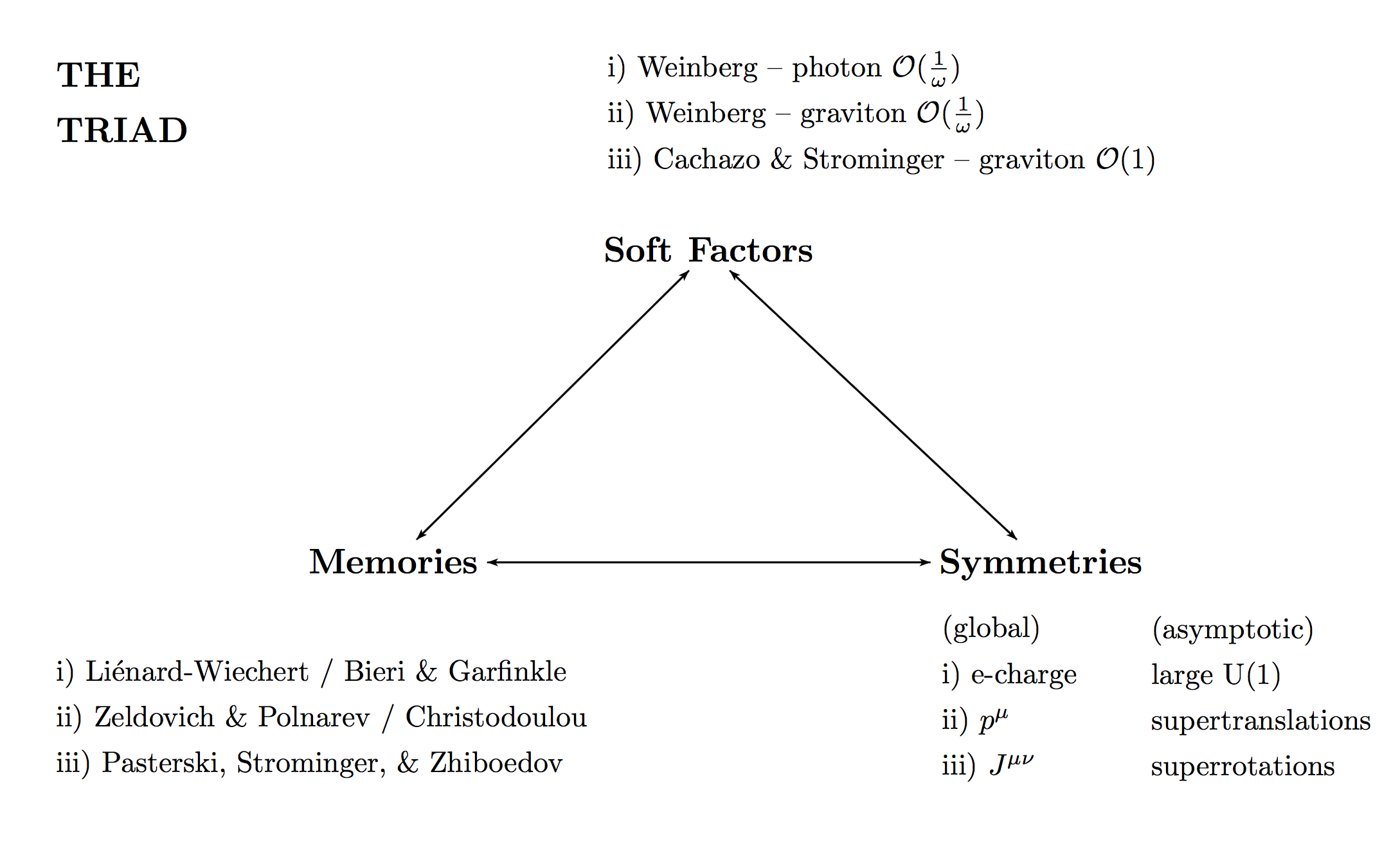}
  \end{center}
\end{figure}

Recent literature has drawn the links connecting soft factors, symmetries, and memories for two of the three sets above.  Of these connections, the oldest and most well known are those that lie between the leading gauge and gravity soft factors and their corresponding global symmetries: charge and four-momentum conservation, respectively, as derived by Weinberg~\cite{steve}.  Also in the 1960's, Bondi, van der Burg, Metzner, and Sachs (BMS) worked out the symmetry group for asymptotically flat spacetimes~\cite{bms}.  In the early 2000's, it was suggested~\cite{Banks:2003vp} that the globally defined BMS  supertranslations could be accompanied by locally defined superrotations, extending the standard homogenous Lorentz group~\cite{bt}.  The utility of such an extension was demonstrated by Strominger and collaborators over the past year when they derived the corresponding tree-level  subleading  soft factor~\cite{fc}, showed its connection to superrotation generators~\cite{klps}, and completed the above triangle by proposing the spin memory effect~\cite{psz}.  The first step, linking soft factors and symmetries, was motivated by concurrent success connecting the leading soft factors with supertranslations~\cite{asbms,hms} and an asymptotic large $U(1)$ gauge symmetry~\cite{hmps}.  The final step of connecting these soft-factors/asymptotic symmetries to a classical observable came in~\cite{sz}, which found that Weingberg's soft graviton theorem corresponds to the gravitational memory effect~\cite{zp,bragthorne, Christodoulou:1991cr}, inspiring the search for and identification of the spin memory effect.

What remains is to draw the final link to the electromagnetic version of a ``memory effect."  We are aided by recent work discussing the electromagnetic analog of gravitational memory~\cite{bg}.  The goal of this paper is to solidify the connection of electromagnetic memory to the asymptotic $U(1)$ gauge symmetry of~\cite{hmps} and the leading Weinberg soft factor.

This paper is organized as follows.  In section~\ref{sec:memo} we clarify what one means by a ``memory" effect, introduce conventions, and set the groundwork for the finite-$r$ measurement interpretation. Section~\ref{sec:mw} describes different manifestations of the electromagnetic memory effect related to the massive/massless splitting of~\cite{bg}. In~\ref{sec:bou}, we outline the applicable boundary conditions. In~\ref{sec:emm}, we discuss equations  relevant to the results of~\cite{bg}.  Section~\ref{sec:wsf} explores the connection to Weinberg's soft factor in the massive case, as can be seen from using retarded radiation solutions in classical electromagnetism a la~\cite{sgp}.  Then in~\ref{sec:u1}, we review the asymptotic $U(1)$ gauge symmetry of~\cite{hmps} and how the previous discussions connect to the new boundary conditions for a massless scattering process.  Finally, section~\ref{sec:discuss} describes an alternative measurement for the electromagnetic memory effect, where suspension of test charges in a viscous fluid results in a net displacement, rather than a velocity kick~\cite{bg},  and concludes the discussion of electromagnetic memory's connections and consequence. 
 
 \section{Memories}\label{sec:memo}
It is useful to clarify what criterion we associate to/use to distinguish a classical observable that we call a ``memory."  The term stems from the gravitational memory effect (see~\cite{sz} for a review), where an array of test masses receive a finite nudge in position as a result of radiation.  Given a scattering process, solving the linearized Einstein equations for the metric perturbation gives a net change in distance.  Gravitational waves (e.g. from an inspiraling binary system) can themselves source such a perturbation in the metric.  One often hears this referred to as the ``non-linear" Christodoulou effect; however,  the same equations can be used to calculate the shift after including the  gravitational contribution to the stress tensor (see the constraint equations in~\cite{psz}).

The essence of this process and its measurement is a ``net effect," (i.e. it probes the zero-frequency limit of the gauge field sourcing the radiation).  This picking out of zero-frequency modes comes from time integration.  In gauge theory and gravity one can construct specific time integrated quantities determined by the same variables used to define $|in\rangle$ and $|out\rangle$ states, making it possible to connect them to $\cs$-matrix Ward identities.  Meanwhile, the fact that these $\cs$-matrix related quantities pick out the zero-frequency modes of the corresponding gauge field motivates why they are connected to soft factors.  The key ingredient to linking these phenomena is the ability to transition between position and momentum space.  

We will now summarize our conventions to make this more precise.  In all three iterations of the symmetry/soft factor/memory triangle, computations are best performed in retarded and advanced coordinates.  The flat Minkowski metric is:  
\be
\begin{array}{lll}
ds^2&=-du^2-2dudr+2r^2\gamma_{z\bz}dzd\bz&~~~(u=t-r)\\
&=-dv^2+2dvdr+2r^2\gamma_{z\bz}dzd\bz&~~~(v=t+r)\\
\end{array}
\ee
in retarded $(u)$ and advanced $(v)$ coordinates, where $\gamma_{z\bz}=\frac{2}{(1+z\bz)^2}$ is the round metric on the $S^2$, with $(z,\bz)$ coordinates describing the stereographic projection of the Riemann sphere
\be
\hat{x}=\frac{1}{1+z\bz}(z+\bz,i(\bz-z),1-z\bz).
\ee
The four-momentum of an on-shell massless particle can thus be parameterized by an energy $(\omega)$ and a direction on the $S^2$:
\be\label{eq:momentum}
q^\mu=\omega(1,\hat{q}).
\ee

\begin{figure}[ht!]
\begin{center}
\includegraphics[width=9cm,clip=true,trim=350 370 200 440]{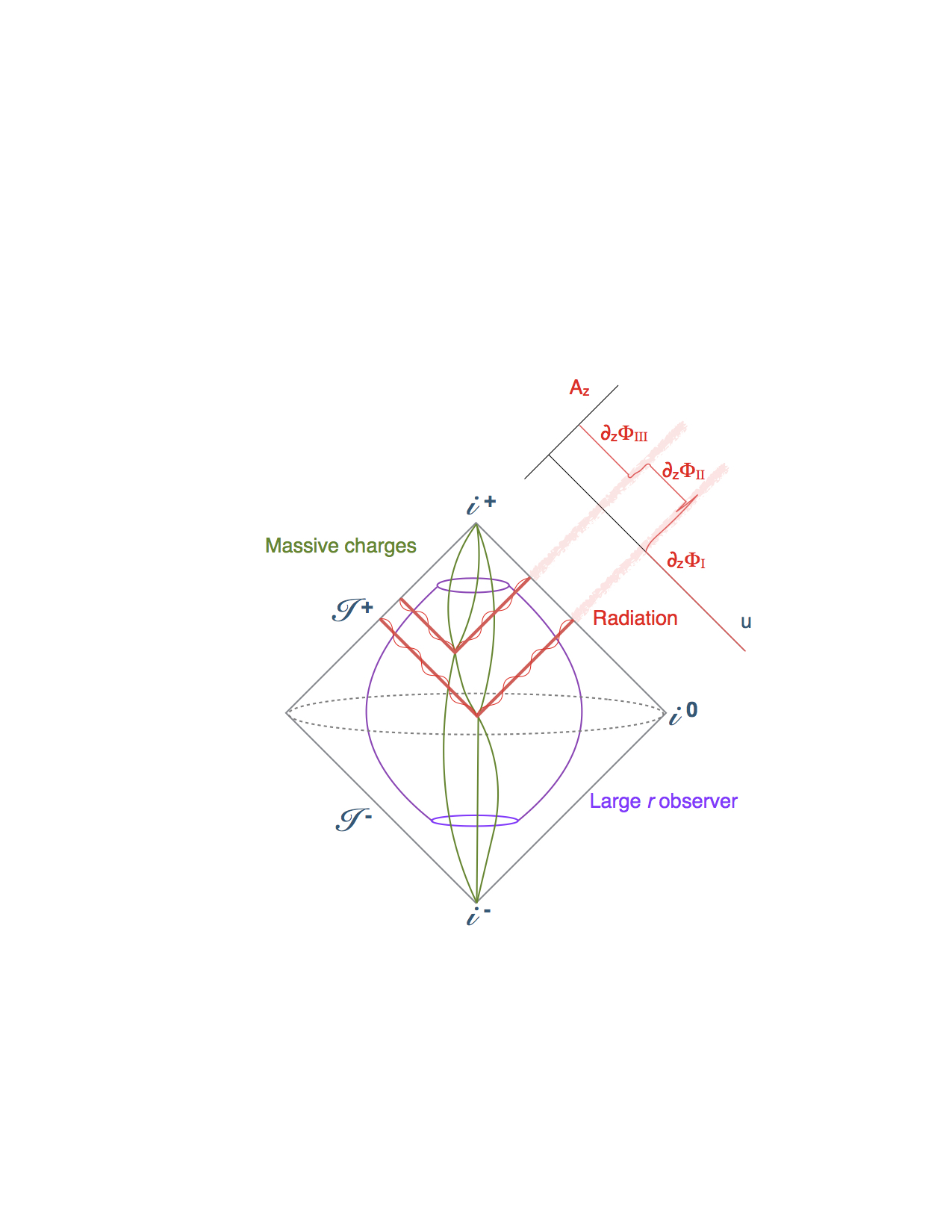}
  \caption{Radiation resulting from the acceleration of charges.  }
  \label{fig:figure}
  \end{center}
\end{figure}

\pagebreak
On a Penrose diagram, massive particles enter at past timelike infinity $i^-$  and exit at future timelike infinity $i^+$, while massless particles enter at past null infinity $\mathcal{I}^-$ and exit at future null infinity $\mathcal{I}^+$ (see Figure~\ref{fig:figure}).  
Making the connection between position and momentum space then relies on the saddle point approximation picking out $\hat{q}\cdot\hat{x}=1$ in the Fourier transformation of the massless field.  Taking retarded coordinates as an example:
\be
e^{iq\cdot x}=e^{-i\omega u-i\omega r(1-\hat{q}\cdot\hat{x})},
\ee
one sees that having the integral over on-shell momenta pick out the parallel direction comes from the order of limits, $r\rightarrow\infty$ first (i.e. before taking $|u|$ large).  

Thinking of quantities such as the asymptotic gauge fields or metric as living on the $\mathbb{R}\times S^2$ of future or past null infinity allows one to separate out the massless from the massive degrees of freedom.  However, when computing quantities that live on null infinity, there should be a way to pull the physical observables into the bulk and make statements at large-but-finite $r$ and also for massive detectors (the generators along $\mathcal{I}$ are null). 

Noting that at fixed-$r$ an integral over all time $t$ becomes an integral along $v$ and then $u$, one can set up a large sphere in the ``radiation zone." (i.e. Accelerating charges/masses sourcing the radiation are assumed to be at a small distance from the center of the sphere compared to the radius $|r_s|<<|r|$.)  Also, when one ``integrates over all time" the relevant changes in the gauge field should start after an early enough $v$ and stop at some late enough $u$.  If one thinks of the measurement sphere as extruding a cylinder ($\mathbb{R}\times S^2$) in spacetime, all massless matter fluxes through the walls of the cylinder while massive matter goes through the endcaps (purple outline in Figure~\ref{fig:figure}).  One must take this into account since states where the particles are moving with constant velocities at early and late times will eventually cross the sphere at some point, but a massive particle will never reach $\mathcal{I}^+$.  The time interval starts and stops the clock when the massive particles are well within the sphere.  Detectors sitting on the sphere in that interval still capture all of the radiation.   (Equivalently, one could restrict oneself to detectors at a large angular separation from where the particles enter or emerge to maintain the radiative $\frac{1}{r}$ power counting.)

The saddle point has the following implication:  the Weinberg soft factor in electromagnetism corresponds to the time integral of the radiated electric field where one replaces $\hat{q}$, describing the direction of the emitted soft photon, with the $\hat{x}$ for the position of the large-$r$ observer measuring the radiation.

 \section{Maxwell in the Radiation Zone}
 \label{sec:mw}
 \subsection{Boundary Conditions}
 \label{sec:bou}

There are two sets of boundary conditions relevant to discussing asymptotic symmetries and the electromagnetic memory effect.  First, specifying the radial fall-off conditions on the electromagnetic fields allows one to solve for the radiation-zone solution to Maxwell's equations.  Second, placing matching conditions on the gauge potential across spatial infinity $i^0$, and adding field strength boundary conditions at the temporal extremes of past and future null infinity, allows one to establish $\cs$-matrix symmetries.  There is more flexibility in the second step.  Multiple methods can consistently give a ``memory effect" with varying degrees of utility as an asymptotic $\cs$-matrix symmetry.

The derivations of the relevant classical field equations in~\cite{hmps} and~\cite{bg} are equivalent with respect to the first of these two steps.  The fall-off conditions include:  i) an $\mathcal{O}(\frac{1}{r^2})$ radial electric field, ii) $\mathcal{O}(\frac{1}{r})$ radiative fields in Cartesian coordinates, and iii) vanishing radial magnetic field (at each angle) at very early and very late times.

The second step of boundary matching will be considered in section~\ref{sec:u1}.  The fundamental choice one confronts is whether to choose only retarded radiation solution or some admixture of advanced and retarded radiation solutions to solve Maxwell's equations.  The underlying question is whether to consider the charges taking part in a scattering process as transmitters or receivers.  If one shoots a charged mass down an otherwise straight, rigid, frictionless wire with a kink in it (we can just as well smooth it out to a rounded elbow), then one can imagine that the mechanical forces causing the charge to accelerate as it rounds the bend will also cause it to emit radiation, making the retarded solution the best choice.  On the other hand, one could look at the effects of incoming radiation on a set of charges.  Explicit CPT symmetry ends up preferring a symmetric combination of incoming and outgoing radiation.

\subsection{Electromagnetic Memory}
\label{sec:emm}
As pointed out by~\cite{bg}, the electromagnetic analog of the gravitational memory effect amounts to the time integrated radiated electric field.  Consistent with~\cite{bg} but in the notation of~\cite{hmps}, the relevant Maxwell equation is
\be
\p_u A_u=\p_u(D^zA_z+D^\bz A_\bz)+e^2j_u,
\ee
where $D$ denotes a covariant derivative with respect to the unit $S^2$ and $j_u$ is the $\mathcal{O}(r^{-2})$ term in the electric charge current.  Here, the radial dependence has been stripped by taking the large $r$ limit and performing a radial expansion of $\mathcal{F}=d\mathcal{A}$.  Explicitly, $A_{\mu}(u,z,\bz)$ are the leading coefficients of the $\frac{1}{r}$ expansion of $\mathcal{A}_\mu(r,u,z,\bz)$ with $\mathcal{A}_u=\mathcal{O}(r^{-1})$, $\mathcal{A}_z=\mathcal{O}(1)$.  The gauge choice of~\cite{hmps} gives the following relations to the large~$r$ limit of the field strength tensor:
\be
\begin{array}{ll}
F_{ur}&=A_u\\
F_{z\bz}&=\p_zA_\bz-\p_\bz A_z\\
F_{uz}&=\p_u A_z,\\
\end{array}
\ee
where $\mathcal{F}_{ur}=\mathcal{O}(r^{-2})$, $\mathcal{F}_{z\bz}=\mathcal{O}(1)$, $\mathcal{F}_{uz}=\mathcal{O}(1)$, and $F_{\mu\nu}$ are the corresponding leading coefficients in the radial expansion of $\mathcal{F}_{\mu\nu}$.
In the case where all of the charged matter is massive, the current will be zero at the position of the detector.  Note that $F_{ur}$ corresponds to the radial electric field ($A_u=-e^2r^2E_r$), $F_{z\bz}$ to the radial magnetic field, and $F_{uz}$ to the radiative fields (tangent to the $S^2$).

\pagebreak 

Integrating along $u$ and using $F_{z\bz}=0$ at the boundaries of $\mathcal{I}^+$ (denoted $\mathcal{I}^+_-$ and $\mathcal{I}^+_+$), one gets at each angle, 
\be\label{eq:jequ}
\Delta A_u=2D^z\Delta A_z+e^2 \int du j_u,
\ee
where this equation is accompanied by the restriction that $\Delta A_z=\p_z \phi$ for some function $\phi(z,\bz)$.  When one considers only the retarded radiation solution, integrating along $u$ at fixed $r$ is equivalent to integrating for all times, since there is no incoming radiation before the scattering.  When all of the charges are massive and the $j_u$ term is zero, one finds that the integrated gauge field is related to the change in the radial electric field.  This is the Coulomb term.  The key then is to look at the radial electric field for a constantly moving charge.  

In section~\ref{sec:wsf}, we will show that $\Delta A_u$ and the Weinberg soft factor are precisely the change in radial electric field for given initial and final configurations of boosted charges.

To make this ``memory" effect official, we would like to prescribe a way of measuring this time integrated electric field that entails setting up, waiting for, and then making a final measurement.  (i.e. One wants a way to extract {\it just} the zero-mode effect.)~~\cite{bg} relates this effect to a velocity kick, but section~\ref{sec:discuss} suggests a more contained measurement that suspends the charge in a viscous fluid to turn the electromagnetic memory into a net displacement, keeping the charge near the same spot on the sphere. 

What intrigues us about the electromagnetic memory is its universal dependence on the incoming and outgoing asymptotic states of the charged particles, while being linear in the electric field.  As~\cite{bg} points out, a typical photon detector would measure the electromagnetic energy flux, which is quadratic in the field strength.  Given the initial and final momenta, Maxwell's equations constrain the net time integrated radiated electric field at any given angle.  This corresponds to the electromagnetic memory.  However, one can imagine distributing this radiation over a very slow ramp.  If we tune down the rate at which charges accelerate, we can make the power flux arbitrarily small, while keeping the same value of the net integrated field because the ramp integrates to the same end point but takes longer to get there.  To keep the position of the accelerating charges near the origin $|r_s|<<|r|$ during this ramp, we can simultaneously consider detectors that are further away to maintain the order of limits consistent with $r\rightarrow\infty$ first.

\pagebreak

\subsection{Weinberg Soft Factor}
\label{sec:wsf}
 The simplest way of seeing the connection between the Weinberg soft factor and the above electromagnetic memory effect is to make a few more assumptions about trajectories so one can evaluate $\Delta A_u$ for the Li\'enard-Wiechert solution and show that it is the same as the Weinberg soft factor with $\hat{q}$ for the soft photon replaced by $\hat{x}$ giving the location of the observer.  
 
First, consider the radial electric field of a boosted, but constantly moving charge for $|r_s|<<|r|$ evaluated at $\vec{x}=r\hat{n}$ in terms of the position of the charge at the retarded time:
\be\label{eq:er}
E_r=\frac{Q}{4\pi r^2}\frac{1}{\gamma^2(1-\vec{\beta}\cdot{\hat{n}})^2}.
\ee
Next, note that the electromagnetic soft factor contribution for a massive particle with momentum $p$ is:
\be
S_p^{(0)\pm}=eQ\frac{p\cdot\e^\pm}{p\cdot q},
\ee
where explicitly 
\be
p^\mu=m\gamma(1,\vec{\beta})
\ee
and $\gamma$ with no indices refers to the Lorentz factor.

By additionally assuming that the acceleration occurs over a small window during which the charges do not appreciably move from the center of the observing sphere, one can evaluate $\Delta A_u$ by subtracting the initial from the final $A_u$ of a superposition of constantly moving charges near the same value of $u$.  (Note how this connects to the $r\rightarrow \infty$ order of limits, squeezing the relevant interactions towards the $u=0$ lightcone if one ``zooms out enough."  On a Penrose diagram, such zooming out amounts to changing the length scale that goes into defining the coordinates for the conformal compactification before plotting the trajectories.)

 The saddle point approximation of the radiative mode expansion gives
  \be\label{eq:da}
\Delta A_z=-\frac{e}{4\pi}\hat{\e}^{*+}_{z}\omega S^{(0)+},
\ee
when interpreting the soft factor as giving the expectation value~\cite{sz}.  Here, the full soft factor is the signed sum of outgoing minus incoming charged particle contributions.  Using~(\ref{eq:momentum}) with $\hat{q}=\hat{n}$, we evaluate:
\be\label{eq:ssf}\begin{array}{ll}
-\frac{e}{4\pi}\lim\limits_{\omega\rightarrow0}\omega [D^z\hat{\e}^{*+}_{z}S_p^{(0)+}+D^\bz  \hat{\e}^{*-}_{\bz}S_p^{(0)-}]&=-e^2\frac{Q}{4\pi}\frac{1}{\gamma^2(1-\vec{\beta}\cdot\hat{n})^2},\\
\end{array}
\ee
where $\hat{\e}$ is the $r$-stripped polarization tensor in retarded radial coordinates.  This connects the single particle contribution to the soft factor to the radial electric field of the asymptotic configuration. Using $A_u=-e^2r^2E_r$, the contributions from (\ref{eq:ssf}) agrees with (\ref{eq:jequ}).

Having early and late asymptotic states with constant on-shell velocities implies this $\Delta A_u$ corresponds to the electromagnetic soft factor.  Consistency of Maxwell's equations at $\mathcal{I}^+$ given a scattering process with no charges exiting $\mathcal{I}^+$, demands the outgoing radiation solution have a net impulse corresponding to the soft factor.\footnote{As a side note, the same analysis can be applied to the leading Weinberg pole in the gravity case.  There the analog of the radial electric field is the boosted Bondi mass $m_B$ in Bondi gauge.  For massive scattering with no flux through $\mathcal{I}^+$ the linearized constraint equation and soft factor/expectation value interpretation give:
\be
\Delta m_B=\frac{1}{4}\left[D^zD^z\Delta C_{\bz\bz}+D^zD^z\Delta C_{\bz\bz}\right],~~~\Delta C_{zz}=-\frac{\kappa}{4\pi} \hat{\e}^{*+}_{zz}S^{(0)+}.
\ee 
This is consistent with the analog of (\ref{eq:ssf}):
\be\begin{array}{ll}
-\frac{\kappa}{4\pi}\lim\limits_{\omega\rightarrow0}\omega [D^z D^z \hat{\e}^{*+}_{zz}S_p^{(0)+}+D^\bz D^\bz \hat{\e}^{*-}_{\bz\bz}S_p^{(0)-}]&=\frac{4Gm}{\gamma^3(1-\vec{\beta}\cdot\hat{n})^3}=4m_B(\vec{\beta}),\\
\end{array}
\ee
where the second equality can be compared with \cite{bms} for a boosted mass, and the single particle soft factor contribution is now
$S_p^{(0)\pm}=\frac{\kappa}{2}\frac{(p\cdot\e^\pm)^2}{p\cdot q}$ with $\kappa=\sqrt{32\pi G}$.}

\subsection{Large $U(1)$ Symmetry}
\label{sec:u1}
In this section, we review the derivation of the asymptotic $U(1)$ gauge symmetry found in~\cite{hmps} and discuss the second step in setting the boundary conditions for an $\cs$-matrix symmetry.  

As a primer, let us take a moment to consider how a residual large gauge symmetry can be seen as necessary for self consistency of the theory with radiation along $\mathcal{I}^+$.  Consider the plot in the upper righthand corner of in Figure~\ref{fig:figure}.  One can look at the gauge field at a particular angle on the $S^2$ as a function of $u$.  The first round of boundary conditions resulted in the electromagnetic memory depending on a ``pure gauge" function $\Delta A_{z}=\p_z\phi$.  Consider situations where the durations over which accelerations emitting radiation occur have compact support along $u$.  Then separate out intervals between scattering processes. This follows naturally from assuming one can isolate a single interaction.  One should be able to measure the radiation over the time interval relevant to a particular process and extract information that does not depend on later processes.  As such, one can imagine intervals of ``pure gauge" between each such segment for well-separated events.  Indeed, in the Light-Shell Effective Theory (LSET) solutions for massless scattering considered by~\cite{gks}, consistency with the soft factor comes from a step function profile in the radiation (on the $u=0$ shell propagating from an interaction at the spacetime origin). 

\pagebreak

Whereas the vacuum ``picks out" a starting $A_z$ configuration, if one tries to ``set" the vacuum gauge field at early times to zero, one finds that Weinberg's soft theorem implies that the late time vacuum will {\it generically not be $\phi=0$}.  Moreover, given the picture of well-separated events, between any two wavefronts of radiation (i.e. in any of the three regions in Figure~\ref{fig:figure}), we should be able to ``reset" our baseline (i.e. perform a large gauge transformation to set $A_z=0$ over that interval).  Thus, while one zero mode corresponds to a step, it must naturally be accompanied by an overall shift at each angle which corresponds to the resetting.  Note that one can heuristically see how the presence of the Weinberg pole corresponds to a step, from the fact that the Fourier transformation of a step function is a pole.  Meanwhile, the Fourier transformation of a constant is a delta function, so this overall shift can be added in by hand to the standard Fourier transform modes~\cite{sgp} as a strictly zero frequency extension of the quantum mechanical phase space. 

The fact that any vacuum spontaneously breaks the symmetry in choice of $A_z$ is the origin of the Goldstone mode interpretation of $\phi$.  Among the intriguing aspects of the $\cs$-matrix approach of~\cite{hmps} is the way in which the Ward identity motivates a bracket between these two zero modes by considering the charge generating the large $U(1)$ symmetry (whereas one would have trouble starting from the Hamiltonian, since the pure gauge piece does not contribute to the energy and one does not ``evolve" these time averaged/integrated quantities).

Now to address the second step in boundary matching.  The equations of motion (\ref{eq:jequ}) and the analogous one on $\mathcal{I}^-$ are promoted to operator statements when inserted into the $\cs$-matrix.  In constructing a Ward identity of the form
\be
\langle out|(Q_\e^+\cs-\cs Q^-_\e)|in\rangle=0
\ee
boundary conditions setting $A_u=0$ at $\mathcal{I}^+_+$ and $\mathcal{I}^-_-$, followed by antipodally matching $A_z$ at $\mathcal{I}^+_-$ and $\mathcal{I}^-_+$ across $i^0$ allow one to relate the current to the soft factor.  This is a particular feature of massless scattering, where all the charges enter $\mathcal{I}^-$ and exit $\mathcal{I}^+$ and one can cancel the $\Delta A_u$ term between incoming and outgoing scatterers.  (i.e. The massless soft factor obeys a differential equation that localizes on the $S^2$, just as the massless current does.)
 
The required mix of retarded and advanced solutions comes from the way in which the current, as a generator of gauge transformations on the matter fields, only acts on the outgoing or incoming particles for $j_u$ or $j_v$, respectively, whereas an outgoing soft photon attaches to both incoming and outgoing legs (ditto for an incoming soft photon).  As such, a factor of $\frac{1}{2}$ arises from averaging the incoming and outgoing radiation solutions to match the combined current contribution that counts incoming and outgoing particles only once each.

As a final note, we point out a connection to the $\phi=\phi_{m>0}+\phi_{m=0}$ splitting by~\cite{bg} into components sourced by $\Delta A_u$ from massive charges and $e^2\int du j_u$ from the massless current.  The radiation response due to the massive charges results from a change in their kinematic distributions, whereas the response from the massless charges accounts for them exiting the sphere.  Since the analysis of~\cite{bg} considers only the solution near $\mathcal{I}^+$, one should use crossing symmetry to consider a neutral incoming state to connect with the analysis of~\cite{hmps}.  (For the purpose of visualizing radiation arising from prescribed accelerations of scattered charges, one can imagine  superimposing the non-radiating solution of an oppositely charged particle moving unperturbed through spacetime parallel to each incoming particle that scatters.  While there is no incoming charge, an oppositely charged particle leaving at the antipodal angle maintains the outgoing-minus-incoming structure of the original soft factor).  Between the results shown here and in~\cite{hmps}, the soft factor is consistent with measuring the massless and massive contributions to the memory effect, independently.  This separation provides insight into extending the~\cite{hmps} formalism to the massive case.

~\\

\section{Discussion}
\label{sec:discuss}
Now that we have circuited the triad of connections relevant to the electromagnetic memory effect, let's consider a compact set up that could measure it.  The proposition of~\cite{bg} was to connect the time integrated electric field to a net velocity kick.  Explicitly, a test charge obeying $\vec{F}=m\vec{a}=Q\vec{E}_{rad}$ has an acceleration proportional to $\frac{Q}{m}$ times the radiated electric field (which one should keep in mind is a $\frac{1}{r}$ effect).  If the pulse occurs over a short enough period of time that the test charge remains localized on the sphere, then it receives a net kick in its velocity $\Delta\vec{v}=\int dt \vec{a}$.  

If we prefer to keep the test charge localized rather than letting it fly off at some velocity that would need to be measured (or, if restricted to the sphere, letting it move enough that a path integral of the tangential force would be required), we instead can imagine that at the location of where we want to measure the effect, we have a charged bead suspended in a viscous fluid.  Rather than going too deep into how to realistically separate the scales of the interactions which govern the viscous forces between the bead and the fluid and the scattering-sourced radiation we want to measure, we can imagine an idealized situation where the viscous force dominates and any response to a driving force is proportional to the velocity (i.e. heavily damped rather than inertial).  With a drag force at low Reynolds number of $\vec{F}_D=-\sigma \vec{v}$ for some positive constant $\sigma$, which dominates and balances the driving force from the radiated electric field, one finds:
\be
\int dt Q \vec{E}_{rad}=\int dt \sigma \vec{v}=\sigma \Delta \vec{x}
\ee
in this limit, so that the electromagnetic memory is turned into a net displacement (like in the gravitational memory case) rather than a velocity kick.  To distinguish this effect, the relevant scattering process would need to induce a $\Delta\vec{x}$ larger than the expected drift of the test charge during the integration time, due to Brownian motion.

In summary, we have seen that the connection between asymptotic symmetries, soft factors, and memory effects extends naturally to the $U(1)$ case and rounds out the interpretation of any individual link or vertex in this triad.  Memory effects pick out zero-mode classical observables.  Meanwhile, the position space interpretation of soft factors connects large distances with low frequency radiation in the same direction.  In this manner, soft factors can both: i) lead Ward identities that validate the quantum versions of these symmetries, and ii) give the expectation value of classical radiation measurements.  Furthermore, the ability to superimpose classical radiation solutions corresponding to the memory effect for separate scattering processes, combined with the freedom to reset the gauge field to pure vacuum between pulses of radiation when performing calculations, illustrates from a semi-classical perspective how the presence of ``pure gauge" zero modes are essential for self consistency and should be included in the extended phase space.

~\\
~\\
~\\

\begin{figure}[ht!]
\begin{center}
\includegraphics[width=9cm,clip=true,trim=20 0 20 0]{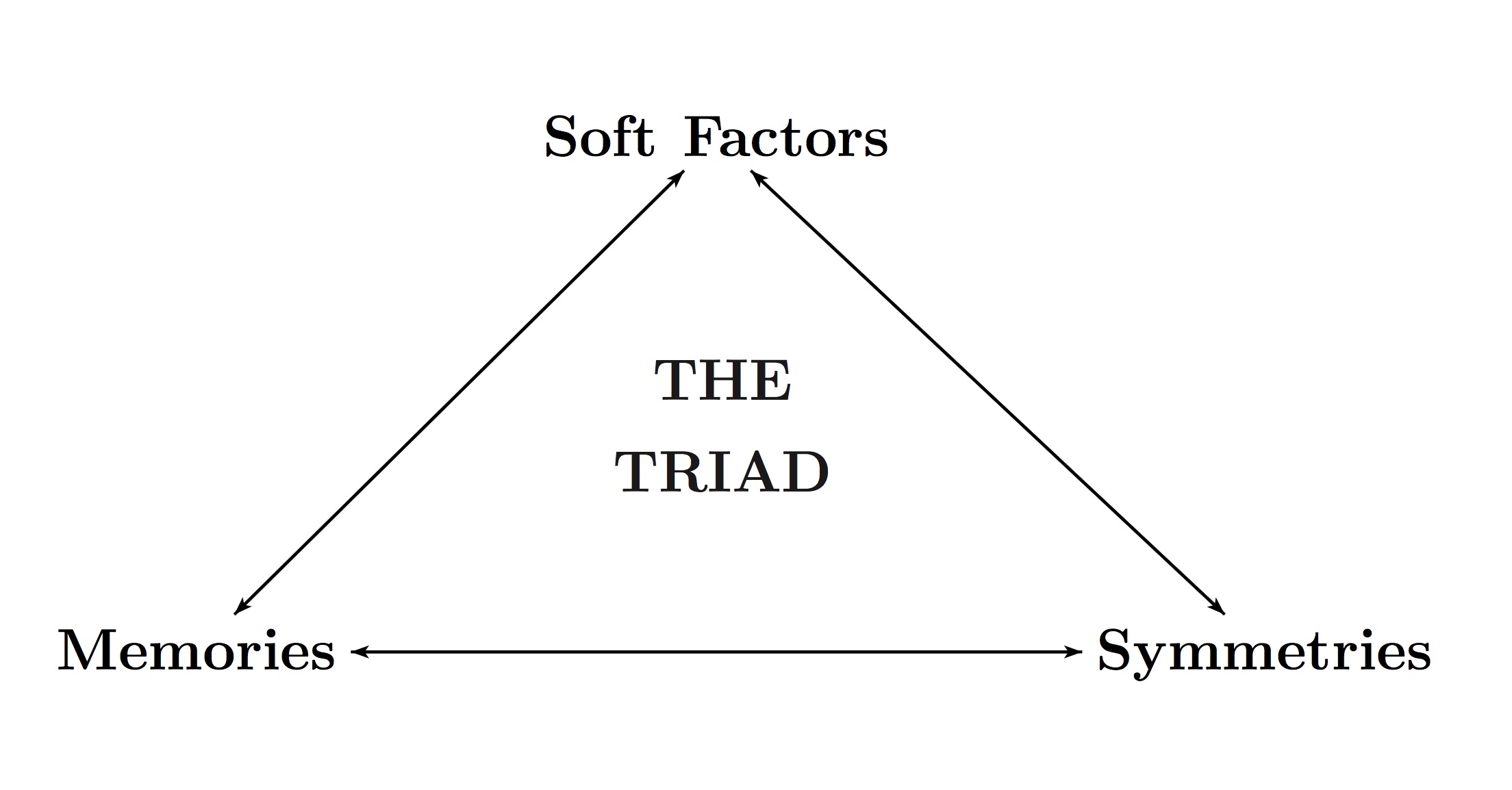}
   \end{center}
\end{figure}

\section*{Acknowledgements}
Many thanks to J. Barandes and A. Zhiboedov.  Thank you to G. Comp\`ere, M. Schwartz, and A. Strominger  for useful questions.  This work coalesced while preparing for Harvard String Family and MIT LHC/BSM Journal Club talks.  I am grateful to the LHC/BSM Journal Club and L. Susskind for convincing me to arXiv my notes.  This work was supported in part by the Fundamental Laws Initiative at Harvard, the Smith Family Foundation, the National Science Foundation, and the Hertz Foundation.

\end{document}